\documentclass[
superscriptaddress,
twocolumn,
nofootinbib,
 amsmath,amssymb,
 aps,
prd,
]{revtex4-2}

\pdfoutput=1
\usepackage{natbib}
\usepackage{aas_macros}
\usepackage{lipsum}
\usepackage{ulem}
\usepackage{graphicx}% Include figure files
\usepackage{dcolumn}% Align table columns on decimal point
\usepackage{bm}% bold math
\usepackage{hyperref}% add hypertext capabilities
\usepackage{booktabs}
\usepackage{color}
\usepackage{hyperref}
\hypersetup{
    colorlinks=true,
    linkcolor=blue,
    filecolor=blue,      
    urlcolor=blue,
    citecolor=blue,
    }

\newcommand{\be}{\begin{equation}}
\newcommand{\ee}{\end{equation}}
\newcommand{\barr}{\begin{eqnarray}}
\newcommand{\earr}{\end{eqnarray}}
\newcommand{\bal} {\begin{aligned}}
\newcommand{\eal} {\end{aligned}}

\begin{document}

\preprint{APS/123-QED}

\title{A New Probe of the High-z BAO scale: BAO tomography With CMB \texorpdfstring{$\times$}{TEXT} LIM-Nulling Convergence}

\author{Hannah Fronenberg}
\email{hannah.fronenberg@mail.mcgill.ca}
\affiliation{Department of Physics, McGill University, 3600 Rue University, Montreal, QC H3A 2T8, Canada}
\affiliation{Trottier Space Institute, 3550 Rue University, Montreal, QC H3A 2A7, Canada}
\author{Abhishek S. Maniyar}
\affiliation{Department of Physics, Stanford University, 382 Via Pueblo Mall, Stanford, CA 94305, USA}
\affiliation{SLAC National Accelerator Laboratory, 2575 Sand Hill Rd, Menlo Park, CA 94025, USA}
\affiliation{Kavli Institute for Particle Astrophysics and Cosmology, 382 Via Pueblo Mall Stanford, CA  94305-4060, USA}
% \affiliation{Department of Physics, New York University, 726 Broadway, New York, NY, 10003, USA}%
\author{Adrian Liu}
\affiliation{Department of Physics, McGill University, 3600 Rue University, Montreal, QC H3A 2T8, Canada}
\affiliation{Trottier Space Institute, 3550 Rue University, Montreal, QC H3A 2A7, Canada}
\author{Anthony R. Pullen}
\affiliation{Department of Physics, New York University, 726 Broadway, New York, NY, 10003, USA}%
\affiliation{Center for Computational Astrophysics, Flatiron Institute, New York, NY 10010, USA}%

\date{\today}% It is always \today, today,
\begin{abstract}
Standard rulers such as the baryon acoustic oscillation (BAO) scale serve as workhorses for precision tests of cosmology, enabling distance measurements that probe the geometry and expansion history of our Universe. Aside from BAO measurements from the cosmic microwave background (CMB), most standard ruler techniques operate at relatively low redshifts and depend on biased tracers of the matter density field. In a companion paper, we explored the scientific reach of nulling estimators, where CMB lensing convergence maps are cross-correlated with linear combinations of similar maps from line intensity mapping (LIM) to precisely null out the low-redshift contributions to CMB lensing. We showed that nulling estimators can be used to constrain the high redshift matter power spectrum and showed that this spectrum exhibits discernible BAO features. Here we propose using these features as a standard ruler at high redshifts that does not rely on biased tracers. Forecasting such a measurement at $z \sim 5$, we find that next-generation instruments will be able to
constrain the BAO scale to $7.2\%$ precision, while our futuristic observing scenario can constrain the BAO scale to $4\%$ precision. This constitutes a fundamentally new kind of BAO measurement during early epochs in our cosmic history.

\end{abstract}

\maketitle

%\tableofcontents

\textit{Introduction}---At the time of recombination ($z \sim 1100$), the first atoms formed and photons streamed freely through the universe. Today, we observe those photons as the cosmic microwave background (CMB), a map which provides unparalleled insight into the early moments at the surface of last scattering. These photons, however, have not travelled unimpeded. Weak gravitational lensing of the CMB arises when photons from the surface of last scattering are deflected by the gravitational potentials they encounter on their way to us, resulting in distortions to the statistics of the CMB. With the use of lensing estimators, one can reconstruct the gravitational potential of the projected mass distribution along the line of sight (LOS) \cite{Hu_Okamoto_2002,Okamoto_Hu_2003}. Reconstructing the lensing potential, or equivalently the convergence, yields direct constraints on the total matter distribution of the universe, both baryonic and dark, without the use of a biased tracer. As such, the CMB convergence has the ability to probe the growth of matter fluctuations, place limits on primordial non-Gaussianity, constrain the sum of the neutrino masses, and even test theories of modified gravity \cite{Zaldarriaga_1998,Lewis_Challinor2007,Schmittfull_2018,Allison_2015}.

This information, however, is projected onto a single observable and the high-redshift contribution to the convergence is dwarfed by that of the low-redshift universe ($z \lesssim 2$). This places limitations on the inferences that can be made about the matter density field. Most notably, the fact that the convergence is a LOS integrated quantity results in the washing out of baryon acoustic oscillations (BAOs). During the radiation--dominated era, dark matter began to cluster while the photon--baryon fluid continued to oscillate, producing BAOs which were then left imprinted on the CMB at recombination. Through large scale structure formation, the BAO scale remains embossed in the distribution of galaxies, and provides a standard distance measure across cosmic time. 

Luckily, line intensity maps (LIMs) also experience weak lensing by large scale structure. These lines, however, are only lensed by a portion of large scale structure that lenses the CMB, namely the low redshift universe. In  Ref. \cite{Maniyar_nulling}, it has been shown that by using the lensing information of two LIMs, one could not just suppress, but exactly null out the low redshift contribution to the CMB convergence. This nulling method has been explored in the context of galaxy lensing \cite{Huterer_nulling_2005, Bernardeau_nulling_2014,Barthelemy_nulling_2020} and has been shown to be able to remove the imprint of various effects from CMB lensing maps for which the physics is uncertain \cite{McCarthy_Foreman_vanEngelen_2021, Qu_2023, Zhang_Omori_Chang_2022}. While never implemented with real data, these techniques can be an important new tool for studying the high redshift universe. What is more, Ref. \cite{Maniyar_nulling} show that the CMB $\times$ LIM-nulling convergence spectrum, $\langle \hat{\kappa}\hat{\kappa}_{\rm{null}} \rangle$, is free of LIM interloper bias. Most recently, we showed in Ref. \cite{LIM_nulling_Fisher} that the CMB $\times$ LIM-nulling convergence can be compared to traditional CMB lensing constraints, and that such comparisons can serve as model-independent tests of cosmology beyond $\Lambda$CDM. Additionally, we explicitly showed that the CMB $\times$ LIM-nulling convergence probes high-redshift modes of the matter power spectrum, which can in turn be used to place limits on the matter power amplitude, the matter transfer function, and measure important features of the matter power spectrum. These results suggest that performing LIM-nulling of CMB lensing observations has the potential to probe a vast amount of high-redshift information. 

\begin{figure}[h]
\includegraphics[width =8.6cm]{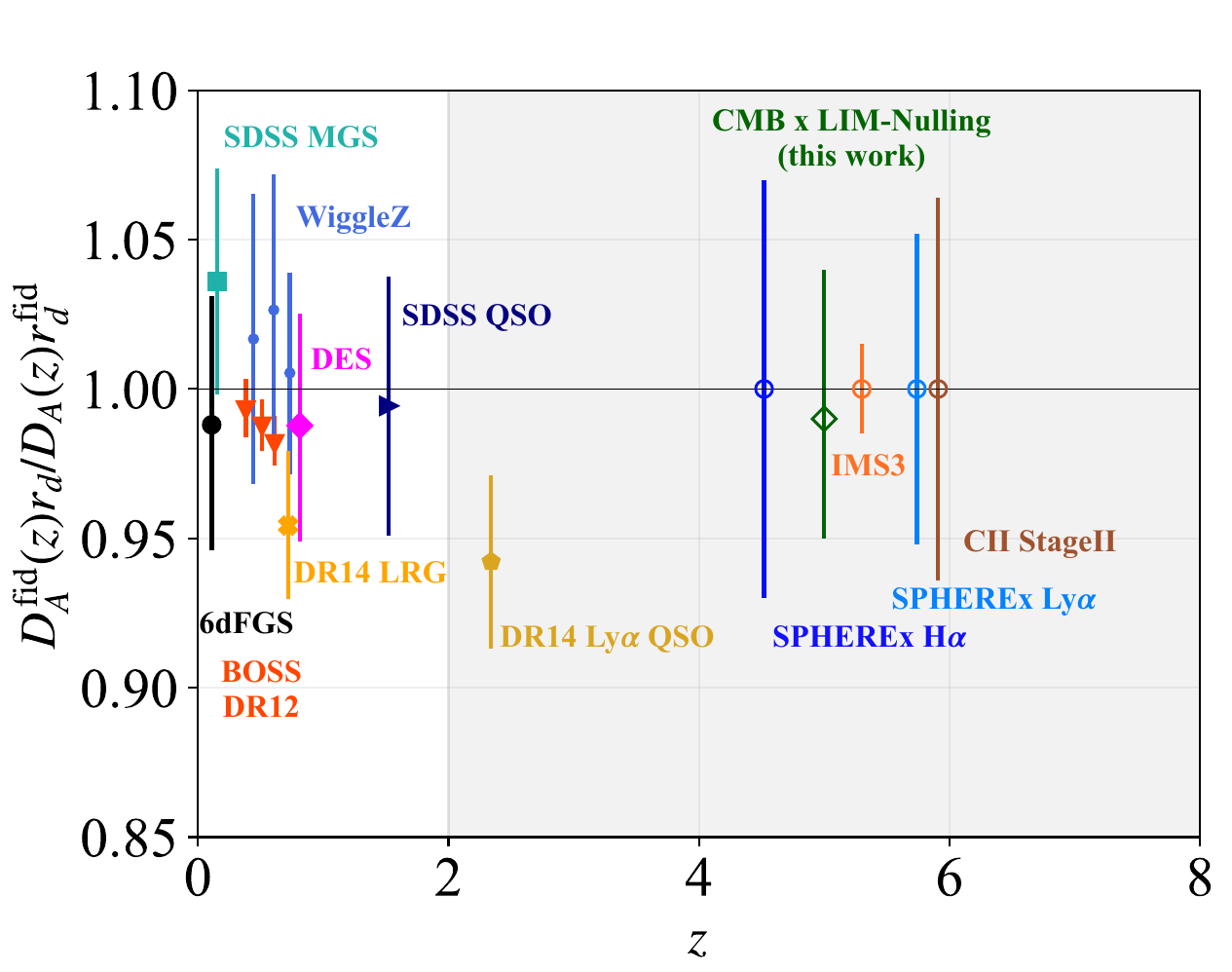}
\caption{Current and projected BAO measurements as a function of redshift. Current measurements from galaxy, quasar, and Lyman-$\alpha$ surveys are shown at $z < 3$ using solid markers \cite{6dFGS,SDSS_DR7,SDSS_III,WiggleZ_BAO,DES_1,SDSS_DR12,SDSS_4,SDSS_3,DES_H0}. Forecast constraints are shown with a hollow marker. The dark green diamond shows the projected measurement from CMB $\times$ LIM-nulling while the remaining constraints are from high-z LIM surveys \cite{Bernal_BAO}. The grey shaded region denotes the redshifts where LIM-nulling can be performed for high-$z$ studies ($2 < z \lesssim 14$).}
\label{fig:BAO_constraints}
\end{figure}

In this \textit{Letter}, we propose a new method to detect BAOs using the CMB $\times$ LIM-nulling convergence spectrum which can be used as a standard ruler over a large cosmological window ($z > 2$). This method, in principle, allows one to directly probe the matter density field during cosmic noon, the epoch of reionization (EoR), cosmic dawn, and even during the cosmic dark ages. The procedure to measure BAOs in this cross-convergence spectrum is similar to that of using the matter power spectrum or the galaxy correlation function; however, this probe has unique benefits. In Fig. \ref{fig:BAO_constraints}, the window in which CMB $\times$ LIM-nulling can be used for studying the high-$z$ universe is shaded in grey.  LIM surveys typically have fine spectral resolution and are conducted over a large bandwidth. This therefore allows for direct, large scale studies of the matter density field over a vast and relatively unexplored cosmic window, filling a void between BAO measurements from $z < 3$ and high-$z$ CMB measurements \cite{6dFGS,SDSS_DR7,SDSS_III,WiggleZ_BAO,DES_1,SDSS_DR12,SDSS_4,SDSS_3,DES_H0}.

\textit{CMB convergence and LIM-nulling}---The CMB acts as a source image which is lensed by the intervening matter density field. The deflection of CMB photons produces correlations between the otherwise uncorrelated CMB spherical harmonic coefficients, $a_{\ell m}$.  Just like the CMB, lower redshift LIMs also incur such correlation as a result of lensing and, most importantly, LIMs and the CMB share common low redshift induced correlations. With the use of quadratic estimators (or more advanced techniques \cite{2003PhRvD..68h3002H,2003PhRvD..67d3001H}), the lensing convergence can be estimated. This estimated quantity is related to the total matter density integrated along the line of sight. The convergence $\kappa$ is given by 
\begin{equation}\label{eq:convergence}
    \kappa(\hat{n}) = \int_0^{z_{\rm{s}}} W(z',z_{\rm{s}})\delta_{m} (\chi(z')\hat{n},z')\frac{c \, dz'}{H(z')}
\end{equation}
where $z_{\rm{s}}$ is the redshift of the source, $ W(z,z_s)$ is the lensing kernel, $H(z)$ is the Hubble parameter, $c$ is the speed of light, $\chi$ denotes the comoving distance, and $\delta_{m} (\mathbf{r},z)$ is the matter overdensity field. The lensing kernel for a source at a single comoving slice is given by
\begin{equation}\label{eq:lensing_kernel}
    W(z,z_s) = \frac{3}{2}\left(\frac{H_0}{c}\right)^2\frac{\Omega_{m,0}}{a}\chi(z)\left(1-\frac{\chi(z)}{\chi(z_s)}\right),
\end{equation}
where $H_0$ is the Hubble constant, $\Omega_{m,0}$ is the matter fraction today, $a$ is the scale factor, and $c$ is the speed of light, and $\chi(z_s)$ is the comoving distance to the source. One can then compute the angular power spectrum $C_{L}^{\kappa_i \kappa_j}$ of any two convergence maps $\kappa_i$ and $\kappa_j$, given by 
\begin{multline}\label{eq:CMB_spectrum}
    C_{L}^{\kappa_i \kappa_j} = \int_0^{z_s} \frac{W_i(z',z_{\rm{s}})W_j(z',z_{\rm{s}})}{\chi(z')^2}
    \\ \times P_{\rm{m}}\left(k = \frac{L+1/2}{\chi(z')},z'\right)\frac{c \, dz'}{H(z')},
\end{multline}
where $P_m$ is the matter power spectrum in the Limber approximation, and $i$ and $j$ index the maps. These maps might, for example, be constructed from the linear combination of multiple convergence maps from different probes. 
 
From Eq. \eqref{eq:convergence}, it should be clear that it is possible to construct some $W_j(z,z_{\rm{s}})$ that vanishes over the low redshift interval $[0,z_{\rm{null}}]$. Since $W$ is quadratic in $\chi$, a linear combination of three such kernels suffices to find a non-trivial null solution for the coefficients of this polynomial. As shown in Ref. \cite{Maniyar_nulling}, using convergence maps estimated from two LIMs and from the CMB at redshifts $z_1< z_2 <z_{\rm{CMB}}$ respectively, the LIM-nulling kernel is given by

\begin{equation}
\label{eq:nulling_kernel}
        W_{\rm{null}} = W(z,z_{\rm{CMB}}) + \alpha W(z,z_2) - (1+\alpha)W(z,z_1),
\end{equation}
where $\alpha \equiv [1/\chi(z_{\rm{CMB}})- 1/\chi(z_1)]/[1/\chi(z_1) - 1/\chi(z_2)]$.

\begin{figure}[t]
\includegraphics[width = 8.9cm]{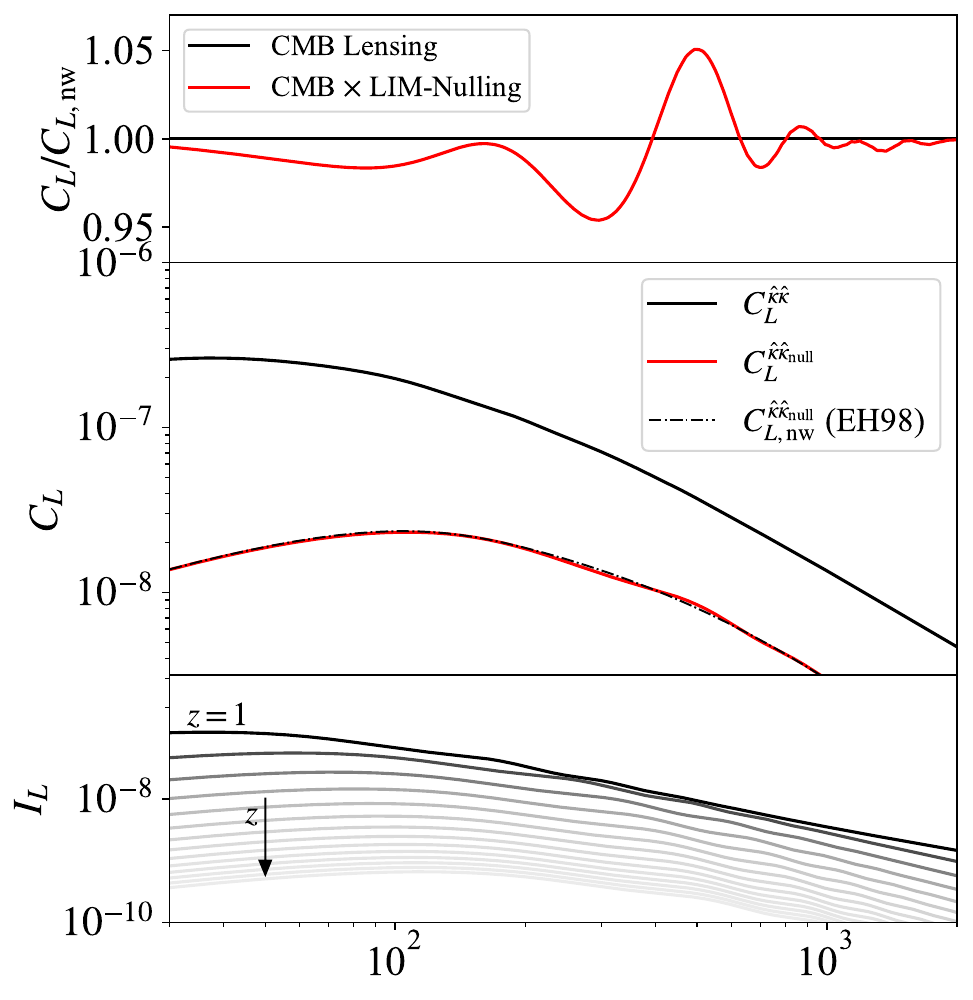}
\caption{Top: Fractional difference between convergence spectra computed with the linear power spectrum and with the no-wiggle Eisenstein \& Hu fitting function for regular CMB lensing (black) and CMB $\times$ LIM-nulling (red). Middle: The CMB convergence spectrum $C^{\hat{\kappa}\hat{\kappa}}_{L}$ (solid black) and the CMB $\times$ LIM-nulling convergence spectrum $C^{\hat{\kappa}\hat{\kappa}{\rm{null}}}_{L}$ (solid red). The dot-dashed black line corresponds to the CMB $\times$ LIM-nulling convergence spectrum computed with the no-wiggle Eisenstein \& Hu fitting function. Bottom: The plot shows the integrand of Eq. \eqref{eq:CMB_spectrum} $I_L$ at increasing redshifts from top (z = 1) to bottom.}
\label{fig:Cl_and_Il}
\end{figure}

Fig. \ref{fig:Cl_and_Il} shows the relevant estimated convergence spectra. The CMB convergence spectrum, $C_{L}^{\hat{\kappa}\hat{\kappa}}$, is smooth with no discernible BAO features which is due to the angular evolution of the BAO wiggles which is depicted in the bottom panel. Examining the integrand of Eq. \ref{eq:CMB_spectrum} evaluated at several redshifts, one can see that the BAO wiggles evolve to lower $L$ as $z$ decreases which, when integrated over, results in the washing out of BAO wiggles in $C_{L}^{\hat{\kappa}\hat{\kappa}}$. While $C_{L}^{\hat{\kappa}\hat{\kappa}}$ can provide information about the scale of structures which dominate the deflection of CMB photons, it cannot act as a standard ruler.

In the CMB $\times$ LIM-nulled convergence spectrum, $C_{L}^{\hat{\kappa}\hat{\kappa}_{\rm{null}}}$, acoustic peaks emerge which is apparent in the top panel of Fig. \ref{fig:Cl_and_Il} where the fractional difference between the wiggle and no-wiggle spectra are plotted. This BAO feature is the result of the much slower angular evolution of the BAO scale at early times. Again, referring now to the bottom panel of Fig. \ref{fig:Cl_and_Il}, the pale high-$z$ curves share largely overlapping acoustic peaks compared to the darker low-$z$ curves whose peaks and troughs mix.

What we argue is that the acoustic peak in the CMB $\times$ LIM-nulling spectrum is, to good approximation, a proxy for measuring the BAO scale at $z = z_{\rm{null}}$. Since the nulling kernel is sharply peaked near $z_{\rm{null}}$ and the matter transfer function is monotonically increasing, the matter density field is weighted most heavily near this redshift. Therefore, the location of the BAO wiggles in the CMB $\times$ LIM-nulling spectrum can be used to measure the BAO scale at $z\sim z_{\rm{null}}$. To test this hypothesis, we perform an Alcock-Paczynski (AP) test on a mock data set in order to place constraints on the BAO scale.

\textit{BAO Model and AP Test}---To model the  CMB $\times$ LIM-nulling convergence spectrum, one can write down the spectrum using Eq. \eqref{eq:CMB_spectrum} with $ i = $ CMB and $j = $LIM-nulling, and parameterise the matter power spectrum using $P_{\rm{model}}(k,z) = P_{\rm{nw}}(k,z) + AP_{\rm{BAO}}(k' = \omega k,z)$ where the BAO wiggles are independently parameterised by $A$ and $\omega$. These parameters control the amplitude and location of the wiggles respectively. The BAO spectrum is given by $P_{\rm{BAO}} = P_{\rm{lin}}- P_{\rm{nw}}$ where $P_{\rm{lin}}$ and $P_{\rm{nw}}$ are the Eisentsein \& Hu linear and no-wiggle fitting functions computed using the publicly available code \texttt{nbodykit} \cite{EH98,nbodykit}. Using this power spectrum model, we fit our two BAO parameters, $A$ and $\omega$. The parameter $A$ is the amplitude of the BAO wiggles in the matter power spectrum and $\omega$ stretches the position of the wiggles as a function of wave-number $k$. The parameter $\omega$ is of particular interest since a change in the configuration space BAO scale is captured by our dilatation parameter $\omega$.

We perform an AP test on mock CMB $\times$ LIM-nulling convergence spectra in order to make use of the BAO wiggle as a standard ruler. Such a test constitutes altering the location of features at some wave number $k$ to $k/\alpha$. Typically the AP parameter $\alpha$ is decomposed into LOS, $k_{||}/\alpha_{||}$, and perpendicular modes, $k_{\perp}/\alpha_{\perp}$, and these are related to the following physical parameters via the relations 
\noindent\begin{minipage}{.5\linewidth}
\begin{equation*}
    \alpha_{\perp} = \frac{D_{A}^{\rm{fid}}(z)r_d}{D_{A}(z)r^{\rm{fid}}_d}
\end{equation*}
\vspace{0.1in}
\end{minipage}%
\begin{minipage}{.5\linewidth}
\begin{equation}
    \alpha_{||} = \frac{H^{\rm{fid}}(z)r^{\rm{fid}}_d}{H(z)r_d}
\end{equation}
\vspace{0.1in}
\end{minipage}
where $H$ is the Hubble parameter, $D_A$ is the angular diameter distance, $r_d$ is the acoustic scale, and the superscript ``fid" denotes the value of the parameter in the fiducial cosmology. 

When nulling is performed with with a single pair of LIMs, however, we are only sensitive to the perpendicular AP parameter. This is because $C_{L}^{\hat{\kappa}\hat{\kappa}_{\rm{null}}}$ is a LOS integrated quantity. Thus, instead of fitting for both AP parameters, we focus on $\alpha_{\perp}$ since this is precisely our parameter $\omega$. While in this work we focus on a single LIM-nulling pair, it is in principle possible to perform LIM-nulling tomography at several different redshifts, using pairs of frequency channels from two LIM experiments, to access the LOS information and $\alpha_{||}$. Given the high spectral resolution of current and upcoming LIM surveys, which resolve the BAO scale in the LOS direction, one can perform nulling at several redshift, subtract the nulled convergence maps, and build a CMB convergence cube instead of just a single map. We leave the full 3D convergence estimation and analysis for future work.

\textit{Mock Data and MCMC Set-up}---We simulate nulling performed with Lyman-$\alpha$ (Ly-$\alpha$) and ionized carbon ([CII]) observations from $z = 3$ and $z = 5$ respectively. Line interlopers (where other spectral lines may redshift into one's observational bands) can be a serious concern for LIMs. Here we include H-$\alpha$ and CO ($J$ = 4–3) as line interlopers for Ly-$\alpha$ and [CII] respectively. We simulate the auto- and cross- spectra of these lines using the publicly available code \texttt{Halogen}\footnote{\url{https://github.com/EmmanuelSchaan/HaloGen/tree/LIM}} which uses a halo model formalism based on conditional luminosity functions \cite{Schaan_2021_a,Schaan_2021_b}.

We define Next Generation and Futuristic observing scenarios. For the former we simulate noise from Cosmic Dawn Intensity Mapper (CDIM) for Ly-$\alpha$ observations, the Stage II instrument for [CII] observations, and the Simons Observatory (SO) for CMB observations, over a 100 deg$^{2}$ field (the nominal survey area for these LIM experiments) \cite{CDIM, Silva_2015, SO,LIM_nulling_Fisher}. We assume interloper residuals to be at the 5\% level and we perform LIM lensing reconstruction using $\ell_{\rm{LIM, max}} = 10,000$. For the Futurisic scenario, we model noise from CDIM, Stage II and from CMB Stage-4 (CMBS4), over one quarter of the sky $(f_{\rm{sky}} = 0.25)$ \cite{CMB-S4}. In this scenario we assume 1\% interloper residuals and perform LIM lensing reconstruction with $\ell_{\rm{LIM,max}} = 20,000$. For the SO noise, we use $N^{\hat{\kappa}_{\rm{CMB}}}_\ell$ from the SO noise calculator\footnote{\url{https://github.com/simonsobs/so\_noise\_models/tree/master/LAT\_lensing\_noise}} and for CMBS4 the lensing noise is obtained from the CMBS4 Wiki \cite{Sailer_2021}. Throughout, our fiducial cosmology is that of \textit{Planck} 2015 \cite{Planck_parameters}. Additional details can be found in Section IV of Ref. \cite{LIM_nulling_Fisher}.

Using Eq. \eqref{eq:CMB_spectrum}, our power spectrum model, and the line and instrument models discussed above, we generate mock $C_{L}^{\hat{\kappa}\hat{\kappa}_{\rm{null}}}$ data and draw Gaussian random noise consistent with the CMB $\times$ LIM-nulling reconstruction variance, $\rm{var}_{L}^{\hat{\kappa}\hat{\kappa}_{\rm{null}}}$, which we derive in Appendix A of Ref. \cite{LIM_nulling_Fisher}. In both cases, the data samples multipoles $L_{\rm{min}} = 30$ to $ L_{\rm{max}} = 1500$. We define a Gaussian likelihood $\mathcal{L}$
\begin{equation}
    \ln{\mathcal{L}(\lambda_i)} = -\frac{1}{2} \sum_{L} \frac{(C^{\rm{data}}_L - C^{\rm{model}}_{L}(\lambda_i))^2}{\rm{var}^{\rm{data}}_{L}}
\end{equation}
where $C^{\rm{data}}_L$ is the data set used to constrain the model $C^{\rm{model}}_{L}(\lambda_i)$ with parameters $\lambda_{i}=\{A, \alpha_{\perp}\}$ in our case. We sample the likelihood using the Python package \texttt{emcee} \cite{emcee} and impose a prior on $A$ that it be non-negative, $A \geq 0 $. Given that the BAO scale has been measured to the percent level by galaxy and quasar surveys and to the sub-percent level by \textit{Planck}, we place a Gaussian prior on $\alpha_\perp$ with 10\% error which is consistent with current observations yet conservative.

\begin{figure}[t]
\includegraphics[width = 8.6cm]{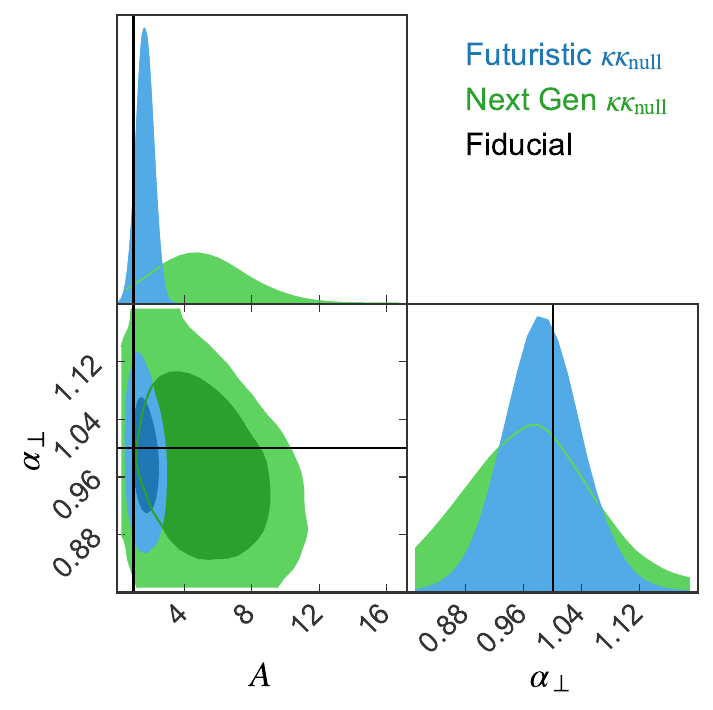}
\caption{Projected 68\% (darker colour) and 95\% (lighter colour) credibility contours for the BAO model parameters $A$ and $\alpha_{\perp}$ for Next Generation (green) and Futuristic (blue) scenarios. The black line denotes the fiducial values.}
\label{fig:Corner}
\end{figure}

\textit{Results}---We present the forecast sensitivities on the model parameters $A$ and $\alpha_{\perp}$ in Table \ref{tab:posteriors} and we show the one- and two-dimensional posterior distributions for these parameters in Fig. \ref{fig:Corner}. In both the Next Generation and Futuristic observing scenarios a BAO detection is possible albeit with relatively low signal-to-noise. The parameter $A$ which characterises the amplitude and therefore the existence of BAO features in the spectrum is weakly constrained in both scenarios, but nonetheless rules out a featureless spectrum at over 3.6$\sigma$ in the optimistic case 1.7$\sigma$ in the moderate case. For the AP parameter $\alpha_{\perp}$, the situation is more encouraging. In the Next Generation scenario, the BAO scale can be measured with to a precision of 7.2\%. In the Futuristic scenario, things are even more promising. It is possible with future generation experiments to measure the BAO scale with 4\% precision solely using CMB lensing information. These constitute BAO scale measurements at $z \sim 5$. While there is still a small amount of decorrelation of BAO wiggles due to the evolution of the comoving distance at $z > 5$, our tests (which involved artificially aligning the BAO features in the bottom panel of Fig.~\ref{fig:Cl_and_Il}) indicate that this is a negligible effect. This test also indicates that any smearing to the effective redshift of this measurement due to the extended kernel is also negligible.

% \usepackage{booktabs}
\begin{table}[h]
\begin{tabular}{@{}llllll@{}}
\toprule
           & \multicolumn{2}{c}{$A$}                        &  & \multicolumn{2}{c}{$\alpha_{\perp}$}           \\ \cmidrule(lr){2-3} \cmidrule(l){5-6} 
           & best fit                   & $\sigma_{\rm{rel}}\%$ &  & best fit                    & $\sigma_{\rm{rel}}\%$ \\ \midrule
Next Gen.    & $5.12^{+3.05}_{-3.09}$ & 59.6                   &  & $0.97^{+0.08}_{-0.07}$ & 7.2              \\
Futuristic & $1.67^{+0.48}_{-0.46}$ & 27.5                  &  & $0.99^{+0.05}_{-0.04}$ & 4.0                 \\ \bottomrule
\end{tabular}
\caption{Posterior on the BAO model parameters for Next Generation and Futuristic cases with 68\% credibility error bars and corresponding relative percent errors.}
\label{tab:posteriors}
\end{table}

To put these results in context, we comment on the current status of BAO measurements from high-redshift LIM surveys. While there are currently no BAO detections at $ 3 < z < 1100 $, there have been a number of forecasts around $z \sim 5$ to which we can compare our results. Ref. \cite{Karkare_BAO_2018} explore the possibility of a BAO detection with CO and [CII] line emission and show that while current generation experiments will be unable to achieve a detection, next-generation and more futuristic experiments could achieve 5\% and $\sim 1\%$ level detection respectively, contingent on the intensity of the line emission. Ref. \cite{Bernal_BAO} show that using SPHEREx H-$\alpha$ measurements at $z = 4.52$ it is possible to measure $\alpha_{\perp}$ to a precision of 7.0\%, the same level as our Next Generation scenario. Using SPHEREx Ly-$\alpha$ they show it is possible to reach a level of precision of 5.2\% at $z = 5.74$. They too include a conception of a futuristic CO mapping instrument which, similar to Ref. \cite{Karkare_BAO_2018}, is able to achieve a 1.5$\%$ level detection at $z = 5.3$. While we do not achieve quite this level of precision in our Futuristic scenario, it should be noted that lensing measurements are not subject to the same astrophysical modelling biases to which LIM BAO measurements are subjected. Since lensing is an unbiased tracer, there is no disentangling the matter power spectrum component from the astrophysical components of the measured spectrum as is the case with LIM surveys.

Considering now BAO forecasts from HI experiments, SKA will not be capable of making BAO detection from the angular direction at frequencies below 800 MHz ($z>0.78$) \cite{Bacon_2020}. This is due to the angular smoothing of BAO features by the SKA beam. SKA will, however, be able to make a BAO detection in the radial direction out to $z \sim 3$.  The detection of velocity-induced acoustic oscillations (VAOs) in the HI power spectrum has also been shown to be a promising standard ruler at cosmic dawn; however, VAOs are damped and undetectable by $z\sim12$ though some work has suggested their signature may persist to lower $z$ \cite{Munoz_2019_PRD, Munoz_2019_PRL,Sarkar_2023,Cain_2020}. Measuring the BAO scale during and soon after reionization is indeed a challenging feat. 

While the constraints we present here are already competitive with existing forecasts, they can still be improved upon. Our forecast here uses just a single pair of LIM frequency channels. By making use of the fine spectral resolution of LIM experiments and combining this spectral information while nulling, it is possible to  yield a higher significance detection. The precise gain in SNR one would obtain from combining spectral information we leave for future work. In addition, the tried and true methods for increasing one's constraints apply here. Increasing the survey areas, performing lensing reconstruction with higher $\ell_{\rm{max}}$, and improving foreground removal techniques for line interlopers can all lower the CMB $\times$ LIM-nulling variance. 
% Moreover, it should be noted that the CMB reconstruction noises are used for this analysis are those estimated using the Hu \& Okamoto lensing estimator. It had been shown that it is possible to yield lower reconstruction noise for CMB lensing measurements using the global-minimum-variance (GMV) estimator \cite{Maniyar_QE} as well as with full likelihood based methods \cite{Hirata_Seljak_2003} which then carries over to yield a lower CMB $\times$ LIM-nulling variance. 
Of course, another avenue would be to place more stringent priors on the fit, although care must be taken to ensure that the posterior is not prior-dominated.

\textit{Conclusion}---We have shown that it is possible to measure the BAO scale over a wide redshift range from a CMB lensing observable. The BAO features that emerge in the CMB $\times$ LIM-nulling convergence spectrum serve as a proxy for the BAO features at $z \sim z_{\rm{null}}$. Using next generation instruments, we have shown, by way of an AP test, that the detection of this feature can act as a standard ruler, and can constrain the perpendicular AP parameter, $\alpha_{\perp}$, to the $4\%$ level in a Futuristic observing scenario and to 7.2$\%$ in the Next Generation scenario. These constrains are competitive with existing high-$z$ forecasts of BAO measurements from LIM surveys. This technique may be used to tease out information about the matter density field over a large period of cosmic history from difficult to reach redshifts purely using information from lensed CMB photons.

\textit{Acknowledgements}--- The authors are delighted to thank Marta Spinelli and Alexandre Refregier for useful discussions and would especially like to thank José Luis Bernal, Gabriela Sato-Polito, Julian Mu\~{n}oz, Simon Foreman, and Eiichiro Komatsu for their insightful comments on an earlier draft. HF is supported by the Fonds de recherche du Québec Nature et Technologies (FRQNT) Doctoral Research Scholarship award number 315907 and acknowledges support from the Mitacs Globalink Research Award for this work. AL and HF acknowledge support from the Trottier Space Institute, the New Frontiers in Research Fund Exploration grant program, the Canadian Institute for Advanced Research (CIFAR) Azrieli Global Scholars program, a Natural Sciences and Engineering Research Council of Canada (NSERC) Discovery Grant and a Discovery Launch Supplement, the Sloan Research Fellowship, and the William Dawson Scholarship at McGill. ARP was supported by NASA under award numbers 80NSSC18K1014, NNH17ZDA001N, and 80NSSC22K0666, and by the NSF under award number 2108411. ARP was also supported by the Simons Foundation. This research was enabled in part by support provided by Compute Canada (\href{www.computecanada.ca}{www.computecanada.ca}).

\bibliographystyle{prsty}

\bibliography{apssamp}% Produces the bibliography via BibTeX.

\end{document}